\documentstyle[12pt,aaspp4]{article}

\begin{document}

   \title{Theoretical models for classical Cepheids. VIII.
Effects of helium and heavy elements abundance on the Cepheid distance scale}

\author{G. Fiorentino \altaffilmark{1}, F. Caputo \altaffilmark{2}, M. Marconi \altaffilmark{1}, I. Musella \altaffilmark{1}}

\affil{1. Osservatorio Astronomico di Capodimonte, Via Moiariello 16,
80131 Napoli, Italy; marcella@na.astro.it; ilaria@na.astro.it}
\affil{2. Osservatorio Astronomico di Roma, Via Frascati 33,
00040 Monte Porzio Catone, Italy; caputo@coma.mporzio.astro.it}

   \begin{abstract}
Previous nonlinear fundamental pulsation models  for classical
Cepheids with metal content $Z\le$ 0.02 are implemented with new
computations at super-solar metallicity ($Z$=0.03, 0.04) and
selected choices of the  helium-to-metal enrichment ratio $\Delta
Y/\Delta Z$. On this basis, we show that the location into the HR
diagram of the Cepheid  instability strip is dependent on both
metal and helium abundance, moving towards higher effective
temperatures with decreasing the metal content (at fixed $Y$) or
with increasing the helium content (at fixed $Z$). The
contributions of helium and metals to the predicted
Period-Luminosity and Period-Luminosity-Color relations are
discussed, as well as the implications on the Cepheid distance
scale. We  suggest that the adoption of empirical $V$ and $I$
Period-Luminosity relations, as inferred by Cepheids at the Large
Magellanic Cloud (LMC), to get distance moduli with an uncertainty
of $\pm$ 0.10 mag is  fully justified  for variables in the short
period range ($P\le$ 10 days), at least with $Z\le$ 0.04 and
$\Delta Y/\Delta Z$ in the range of 2 to 4. Conversely, at longer
periods ($P>$ 10 days) a correction to LMC-based distance
moduli may be needed, whose {\it sign and amount} depend on the
helium and metal content of the Cepheids. Specifically, from
fundamental pulsators with $Z>$ 0.008 we derive that the
correction (in mag)  may be approximated as
$c=-6.03+17.80Y-$2.80log$Z+8.19Y$log$Z$, with a total intrinsic
uncertainty of $\pm$0.05 mag, whereas is
$c=-0.23(\pm$0.03)log$(Z/0.008)$ if $Z<$ 0.008. Based on these new
results, we show that the empirical metallicity correction
suggested by Cepheid observations in two fields of the galaxy M101
may be accounted for, provided that the adopted helium-to-metal
enrichment ratio is reasonably high ($\Delta Y/\Delta Z\sim$ 3.5).

      \keywords{Stars: variables: Cepheids --
               Stars:  oscillations --
               Stars: distances}
   \end{abstract}

%
\pagebreak

\section{Introduction}
Based on  their characteristic
Period-Luminosity (PL) relation, classical Cepheids are
primary indicators to estimate the distance to Local Group galaxies and
to external galaxies with Hubble Space Telescope (HST) observations.
Moreover, through the calibration of secondary distance indicators, they
allow to study even more distant stellar systems, thus providing
fundamental information on the Hubble constant (see Ferrarese et al. 2000).

The PL relation is traditionally assumed to be independent of the
chemical composition (Iben \& Renzini 1984; Freedman \& Madore
1990) and Cepheid distances are generally derived by adopting {\em
universal} multiband PL relations. The slope of these relations is
the one derived for Cepheids at the Large Magellanic Cloud (LMC,
see Madore \& Freedman 1991; Udalski et al. 1999), whereas the
zero-point is referenced to the LMC distance, as obtained with
independent methods (RR Lyrae stars, Red Giant Branch clump,
SN1987A, etc: see, e.g., Freedman 1988; Walker 1999; Udalski et
al. 1999, and references therein) or to calibrating Cepheids in
the Galactic field  (Feast \& Catchpole 1997; Lanoix et al.
1999).

In the last years, we have  deeply investigated the Cepheid
pulsational behavior  through nonlinear, nonlocal and
time-dependent convective pulsational models which take into
account the coupling between  pulsation and convection. With
respect to linear nonadiabatic models (e.g., Chiosi, Wood \&
Capitanio 1993; Saio \&  Gautschy 1998; Alibert et al. 1999), such
a theoretical approach allows improved predictions on the
effective temperature of  both the blue and red edges of the
instability strip. In addition, they provide the amplitude and
morphology of the light-curves all along the pulsation region (see
Bono, Marconi \& Stellingwerf 1999a, 2000; Bono, Castellani  \&
Marconi 2000 and references therein). Using fundamental models
computed with three different chemical compositions ($Y$=0.25,
$Z$=0.004; $Y$=0.25, $Z$=0.008; $Y$=0.28, $Z$=0.02) taken as
representative of Cepheids in the Magellanic Clouds and the
Galaxy, we have already  shown that the predicted
bolometric magnitude of metal-rich variables is, on average,
fainter than that of metal-poor stars with the same period  (Bono
et al. 1999b). Furthermore, we found that both the slope and
zero-point of synthetic PL relations at different wavelengths
depend on the pulsator metallicity, with the amplitude of the
metallicity effect decreasing from visual to near-infrared
magnitudes (Caputo, Marconi \& Musella 2000a). Also the predicted
Period-Luminosity-Color (PLC) relations at the different
wavelengths turned out to  be, in various degrees, metallicity
dependent. As an example, for a given period and $B-V$ color,
metal-rich pulsators are brighter than metal-poor ones, whereas
they are fainter if the $V-K$ color is adopted (Caputo et al. 2000a).

The predicted PL relations in the $V$ and $I$ bands for the three
selected chemical compositions have been applied by Caputo et al.
(2000b) to the  Cepheids observed within two HST surveys, the
``Extragalactic Distance Scale Key Project'' (Freedman et al.
1994, hereafter KP) and the ``Type Ia Supernova Calibration''
(Saha et al. 1994), in order to estimate the metallicity
correction to the published distances. As a fact, both the HST
projects adopt PL relations calibrated on Cepheids at the LMC,
namely on variables with $Z\sim$ 0.008 (see Luck et al. 1998),
whereas the metallicity of the HST observed Cepheids can be
significantly different from that of LMC Cepheids. This in
view of the fact that the range in the [O/H] metallicity
index\footnote{[O/H]=log(O/H)$-$log(O/H)$_{\odot}$, with
log(O/H)$_{\odot}$=$-$3.10.} covered by HII regions in the host
galactic fields is $\sim$ $-$0.7 to  $\sim$ +0.4 (see Ferrarese et
al. 2000; Zaritsky, Kennicutt \& Huchra 1994), while the LMC value
is [O/H]$\sim-$0.40 (Pagel et al. 1978).

Eventually, Caputo et al. (2000b) derived that the Cepheid
intrinsic distance modulus, as inferred by pulsating models,
depends on the adopted metallicity as
$\delta\mu_0/\delta$log$Z\sim +$0.27 mag dex$^{-1}$. Assuming that
the Cepheid metal content scales with the field oxygen abundance,
i.e. adopting $\Delta$[O/H]$\sim \Delta[Z/X]$, yields that the
predicted metallicity correction (in magnitude) to the KP
intrinsic distance modulus is $\Delta\mu_0/\Delta$[O/H]$\sim
-$0.27 mag dex$^{-1}$, where $\Delta$[O/H] is the difference
between the oxygen metallicity of the Cepheid host galaxy whose
distance we wish to determine and the LMC value.

It is worth mentioning that the slope of the various predicted
relations for fundamental pulsators with $Z$=0.008 is in close
agreement with the observed one for Cepheids at the LMC (see also
Caputo, Marconi, \& Musella 2002). In particular, the slope of the
predicted linear PL$_V$ and PL$_I$ relations for Cepheids with
period log$P\le$1.5, as observed in the LMC, is $-2.75\pm0.02$ and
$-2.98\pm0.01$ (see following section), respectively, in very good
agreement with the  values ($-2.76\pm$0.03 and $-2.96\pm$0.02)
inferred from the huge sample of LMC Cepheids in the OGLE-II
catalog (Udalski et al. 1999) and recently adopted in the revised
KP method (Freedman et al. 2001). Moreover, the predicted
metallicity correction presented by Caputo et al. (2000b) provides
a way to solve the apparent discrepancy between Cepheid and maser
distance to the spiral galaxy NGC 4258 (Caputo et al. 2002).

However, in spite of these promising evidences, the disagreement
with earlier empirical corrections made us quite uneasy. In fact,
the analysis of Cepheids in different fields of M31 (Freedman \&
Madore 1990) and M101 (Kennicutt et al. 1998) suggests a
null or even opposite metallicity effect (see also Kochanek 1997;
Sasselov et al. 1997). According to Kennicutt et al. (1998), the
direct application of the LMC-calibrated $V$ and $I$
PL relations to Cepheids in the inner and outer
fields in M101 yields a metallicity correction of
$\Delta\mu_0/\Delta$[O/H]$\sim +$0.24$\pm$0.16 mag dex$^{-1}$.
More recently, the revised
KP method adopts $\Delta\mu_0/\Delta$[O/H]$\sim +$0.20$\pm$0.20
mag dex$^{-1}$
(see Freedman et al. 2001).

Looking for a possible way to explain the discrepancy between
theory and observations, we note that the theoretical metallicity
correction given in Caputo et al. (2000b) refers to the explored
metallicity range 0.004$\le Z \le$0.02 of the pulsating models,
whereas the oxygen abundance of the M101 outer field is close to
the LMC value and that of the inner field is $\sim$ 0.7 dex
larger, suggesting Cepheid metal content of $Z\sim$ 0.008 to
$\sim$ 0.04, respectively. On this basis, we decided to explore
the behavior of very metal rich ($Z>$ 0.02) pulsators with a view
to improving on the predicted metallicity correction to the whole
set of HST galaxies, whose [O/H] index ranges from $\sim-$0.7 to
$\sim+$0.4~dex. Moreover, as the helium abundance of the previous
models was fixed following a primordial helium content $Y_p$=0.23
and a relative helium-to-metal enrichment $\Delta Y/\Delta
Z\sim$~2.5, we decided to analyze the effect of the simultaneous
helium increase that is expected to become important for
super-solar metal abundances.

The organization of the paper is the following. In  Section 2 we present
the pulsation models, together with the results on the predicted instability
strips. The implications for the PL and PLC relations
and the Cepheid distance scale are discussed in Section 3. The
conclusions close the paper.

\section{Pulsation models and theoretical instability strip}

Following the physical and numerical approach described in Bono et
al. (1999a), we have computed new sets of nonlinear convective
models with $Z$=0.03 and $Z$=0.04.  Adopting $Y_p$=0.23 and
$\Delta Y/\Delta Z$=2.5, the corresponding helium abundances are
$Y$=0.31 and $Y$=0.33, respectively. In order to investigate the
effect of a variation of the helium abundance at fixed
metallicity, additional models with $Z$=0.02, $Y$=0.31 and
$Z$=0.04, $Y$=0.39 (corresponding to $\Delta Y/\Delta Z$=4) have
been calculated. For all the computations, the adopted stellar
masses are $M/M_{\odot}=$5, 7, 9, and 11, and for each stellar
mass the luminosity level is fixed following the chemical
composition dependent mass-luminosity (ML) relation provided by
Bono et al. (2000). The new ML relation is slightly different from
the one adopted in our previous papers in the sense that at
fixed mass it predicts slightly lower luminosities ($\sim$ 0.07
dex) with respect to that adopted in our previous investigations.
However, as a decrease in the luminosity reflects in a decrease in
the pulsation period, we evaluate that the final effect on the PL
and PLC relations is of a few hundredths  of a magnitude.
Finally, for each stellar mass, a wide range of effective
temperature has been explored, leading to the predicted boundaries
of the fundamental instability strip, as well as to the bolometric
light curves all along the pulsation region, for each selected
chemical composition. The intrinsic stellar parameters of the
computed models are reported in Tables 1, 2 and 3, together with
the computed periods (in days).

The upper panel in Fig. 1 shows the HR diagram of the predicted
instability strip for fundamental pulsators with $Z$ in the range
of 0.008 to 0.04 and helium abundance, according to $\Delta
Y/\Delta Z$=2.5, from $Y$=0.25 to 0.33, respectively. One has that
as the metal content increases from $Z$=0.008 to 0.03, and
accordingly $Y$ increases from 0.25 to 0.31, both the edges of the
pulsation region move toward lower effective temperatures, leaving
quite unchanged the width of the instability strip. A further
increase from $Z$=0.03 (dashed line) to 0.04 (long-dashed line)
causes
 that
the blue boundary moves towards the red whereas the red boundary
shows an opposite behavior, with the final effect of narrowing the
pulsation region. This last evidence is a result of the decreased
pulsation driving associated to the hydrogen ionization region, as
due to the decreased hydrogen abundance from $X$=0.66 to 0.62. In
fact, since the earlier investigations by Baker \& Kippenhan
(1965), Unno (1965), Christy (1962), Cox et al. (1965) it has become
clear that  the
pulsation driving associated to the H ionization region can be
comparable with the one associated to the second Helium ionization
region. More recently,  Bono et al. (1999a) have
shown that the contribution of the H ionization region to the
pulsation driving of Classical Cepheids is expected to become
dominant in particular regions of the instability strip. The
significant reduction of the H content, as due to the simultaneous
increase of helium and metal abundances, causes a decrease in the
local driving supplied by this zone, reducing the pulsation
efficiency and, in turn,  the width of the instability strip.

The lower panel in Fig. 1 shows the instability strip for
pulsators with $Z$=0.02 and $Y$=0.31 ($\Delta Y/\Delta Z=4$,
dashed-dotted line) in comparison with the results at fixed metal
content ($Z$=0.02 and $Y$=0.28, solid line) and helium abundance
($Z$=0.03 and $Y$=0.31, dashed line). For fixed He abundance, one
has that the effect of a larger $Z$ is to move the instability
strip towards the red. On the contrary, for fixed $Z$, an
increased $Y$ moves the pulsation region towards higher effective
temperatures, mainly at the higher luminosity levels. As a result,
the pulsation region with $Z$=0.02 and $Y$=0.31 turns out to be
much more close to the predicted one for LMC variables ($Z$=0.008,
dotted line). Eventually, the full set of pulsating models
provides clear evidence that the location of the Cepheid
instability strip in the HR diagram {\it depends on both the
helium and metal content of the pulsators}.

One agrees that, if $Z<$ 0.01, a variation of $\Delta Y/\Delta Z$
in the range 2 to 4 does not significantly modify the helium
content, and in turn the expected pulsational behavior, whereas
with $Z\ge$ 0.02 the effect becomes more and more important. On
this subject, it is worth noticing that no pulsation model results
to be unstable with $Z$=0.04 and $Y$=0.39, i.e. $\Delta Y/\Delta
Z=4$, as a consequence of the extreme reduction of the hydrogen
abundance ($X$=0.57). On the other hand, Bono et al. (2000) have
recently shown that the evolution in the HR diagram of
intermediate-mass stars with $Z$=0.04 is strongly dependent on He
abundances. In particular, an increase from $Y$=0.29 to 0.37
causes a decrease in the temperature excursion of the blue-loop,
with the consequent lower probability to cross the Cepheid
instability strip, and/or a reduction in the time spent inside the
instability strip. Thus, independent pulsational and evolutionary
results seem to provide an upper limit of $\Delta Y/\Delta Z\sim$
3.5 for the detection of super metal-rich ($Z$=0.04) Cepheids.

Before closing this section, we wish also mention that the predicted
relations based on pulsating models
are still affected by non negligible
systematic effects. It has been recently shown
(Caputo et al. 2001; Bono, Castellani, \& Marconi 2002) that the
canonical assumptions on both the mass-luminosity relation and the
mixing length parameter should be revised. However, these
variations are expected to  mainly affect the zero points of
theoretical relations, leaving the results about the dependence on
chemical composition almost unchanged.

\section{Implications for the Cepheid distance scale}

On the basis of the results discussed in the previous section, we
can investigate the combined effects of metal and helium
abundances on the Cepheid PL and PLC relations. To this aim, the
bolometric light curves of the computed models were transformed
into the observational bands $BVRIJK$ by means of the atmosphere
models published
by Castelli, Gratton \& Kurucz (1997a,b) and  specifically computed
by Castelli
(private communication) for super-solar chemical compositions.
The transformed light curves were used to
derive the intensity averaged mean magnitudes and colors presented
in Tables 1, 2 and 3.

The predicted multiband PLC relations for all the computed models
with the various chemical compositions are reported in Table 4. As
for the PL relations, which are well known to depend on the
topology of the instability strip and on the distribution of
pulsators within the strip, we did not use the individual models
but we populated the predicted instability strip by adopting the
procedure suggested by Kennicutt et al. (1998) and already used by
Caputo et al. (2000a). In particular, 1000 pulsators were
uniformly distributed from the blue to the red boundary of the
instability strip, with a mass law as given by $dn/dm=m^{-3}$ over
the mass range 5-11$M_{\odot}$ (see Caputo et al. 2000a for
further details).

The resulting synthetic multiband ($BVRIJK$) PL relations are
given in Table 5 and Table 6 (quadratic and linear solutions,
respectively). We have already shown in previous papers (Bono
et al. 1999b; Caputo et al. 2000a) that moving toward the shorter
wavelengths the $M_{\lambda}$-log$P$ distribution of fundamental
pulsators with periods longer than $\sim$ 3 days is much better
represented by a quadratic relation. On the observational side,
some hints have been presented by Sandage \& Tammann (1968) and
Sasselov et al. (1997) who state "Cepheids with $P>$  50 days are
expected (and observed) to deviate from a linear PL relation".
Moreover, we wish to recall that, as discussed in Ferrarese et al.
(2000), the distance modulus of the KP galaxies increases when the
lower period cut-off is moved from 10 to 20-25 days, suggesting
that the adopted LMC-referenced linear PL relations  are
over-estimating the absolute magnitudes of Cepheids in the
long-period range. On the other hand, the almost perfectly linear
PL relations provided by the huge number of Cepheids in the LMC is
consistent with the theoretical results. As discussed by Caputo,
Marconi \& Musella (2002), if only models with periods shorter
than log$P$=1.5 are considerd in the final fit, then reliable
linear solutions are obtained, with the slope of the predicted
linear PL$_V$ and PL$_I$ relations with $Z$=0.008 as given by
$-2.75\pm0.02$ and $-2.98\pm0.01$, respectively, in very good
agreement with
 the observed
values ($-2.76\pm$0.03 and $-2.96\pm$0.02) inferred from the LMC
Cepheids in the OGLE-II catalog (Udalski et al. 1999). For the
above reasons, the quadratic solutions in Table 5 hold for
pulsators with log$P\ge$ 0.5, while the linear solutions in Table
6 should be applied only with log$P\le$ 1.5.

 Figure 2 shows the behavior of the linear PL$_V$ (top panel)
and PL$_I$ (bottom panel)  relations with $\Delta Y/\Delta Z$=2.5 
in comparison with the results at solar metallicity and 
$\Delta Y/\Delta Z$=4.  
Adopting $\Delta{Y}/\Delta{Z}=2.5$, the slope of the 
predicted relations 
decreases,  and the predicted magnitudes at fixed period  
get fainter,  as the metal content increases from 
 $Z$=0.008 (solid line) to 0.02 (dotted
line) and  
0.03 (dashed line). However, as
already noticed for the instability strip location in the HR 
diagram (see bottom
panel in Fig. 1), the PL relations at solar metal abundance 
become steeper when the helium content increases from $Y$=0.28
to 0.31 (long-dashed line), moving towards the
predicted relations for LMC Cepheids ($Z$=0.008, $Y$=0.25). This
is an important point because the difference between the PL
relations of Galactic and LMC Cepheids is a debated issue in the
recent literature. Current results seem to suggest that the
combined metallicity and helium abundance effects have to be taken
into account in order to properly address this problem.

We are now on condition to evaluate the corrections to
LMC-calibrated true distance moduli, by taking into account the
predicted combined effect of metallicity and helium abundance on
the Cepheid distance scale.

A straight estimate can be made by considering our models with the
various chemical compositions as real Cepheids at the fixed
distance $\mu_0$=0 mag. Then, using the predicted  linear
relations with $Z$=0.008 and $Y$=0.25 we determine the value
$\mu^L_{0,0.008}$ for all the pulsators. In view of the fact that
HST observations are in the two bands $V$ and $I$, we concentrate
on these PL relations and, according to the KP method, we adopt
$\mu_V-\mu_I=E(V-I)$ and $A_I$/$E(V-I)$=1.54 from the Cardelli, Clayton
\& Mathis (1989) extinction model. In such a way, we are mimicking the
KP procedure and the derived $\mu^L_{0,0.008}$ values provide
directly the effects of chemical composition on the distance
moduli inferred from LMC-calibrated $V$ and $I$ PL relations.

Figure 3 shows the resulting mean values $\mu^L_{0,0.008}$ as a
function of the metal content for the pulsators with $\Delta
Y$/$\Delta Z$=2.5 (circles) and 4 (triangles). Given the different
slope of the predicted PL relations, the discrepancy between
$\mu^L_{0,0.008}$ and the real value ($\mu_0$=0 mag) depends on
the pulsator period. As shown in the upper panel in Fig. 3,
from 3 to $\sim$ 10 days (open circles), the discrepancy is
quite small, thus supporting the adoption of {\it universal
LMC-referenced PL linear relations within this period range}. For
periods in the range 20-30 days (filled circles), the effects
become significant in the metallicity range of $Z\sim$ 0.013  to
0.032, where the adoption of the $Z$=0.008 linear relations yields
pulsators spuriously brighter  by more than 0.1 mag. In
particular, for pulsators with $Z$= 0.02 and $Y$=0.28 we determine
$\mu^L_{0,0.008}\sim$ 0.20 mag. Finally, the data plotted in the
bottom panel in Fig. 3 show that for very long periods ($P>$ 30
days, open circles) the value of $\mu^L_{0,0.008}$ is larger than
0.1 mag over the range  $Z\sim$ 0.01  to 0.04, reaching $\sim$
0.30 mag with $Z$=0.02 and $Y$=0.28.

However, we have shown in the lower panel in Fig. 1 that an
increased helium abundance causes the $Z$=0.02 pulsators to become
bluer, moving the instability strip towards the predicted location
with $Z$=0.008. As a consequence (see also Fig. 2), the
adoption of the $Z$=0.008 linear relations for pulsators with
$Z$=0.02 and $Y$=0.31 (corresponding to $\Delta Y/\Delta Z$=4)
yields $\mu^L_{0,0.008}\sim$ 0.05, 0.09, and 0.15 mag for the
short (open triangle in the upper panel), long (filled triangle),
and very long period period range (open triangle in the bottom
panel), respectively.

On this basis, we estimate that LMC-calibrated linear PL
relations might be safely adopted, with an uncertainty of $\pm$0.1
mag, for Cepheids in the short period range ($P\le$ 10 days), at
least for $Z\le$ 0.04 and $\Delta Y/\Delta Z$=3$\pm$1. Those
relations might be used also for Cepheids with periods in the
range 20-30 days, on condition that $\Delta Y/\Delta Z$ is high
enough ($>$3).

Unfortunately, current estimates of the $\Delta Y/\Delta Z$ ratio
are still disappointingly uncertain, either in the Milky Way or in
external galaxies. As a matter of fact, from extragalactic HII
regions Lequeux et al. (1979) found a value of 3 (with
$Y_p$=0.23), while Pagel et al. (1992) suggested 4$\pm$1 and
Izotov, Thuan, \& Lipovetsky (1997) determined $\Delta Y/\Delta
Z\sim$ 2 (with $Y_p$=0.24). As for the results from main-sequence
nearby stars, the global estimate is 3$\pm$2 (see Pagel \&
Portinari 2000).  In this context, since the Cepheids
observed in HST galaxies, and in faraway galaxies as well, are
generally in the long period range,  we use the whole set of
pulsating models with log$P>$ 1.0  to estimate the average
correction (in magnitude) to $\mu^L_{0,0.008}$ as a function of $Y$
and $Z$.

With metallicity larger than $Z$=0.008 we get the
following analytical relation
$$c=-6.03+17.80Y-2.80\log Z+8.19Y\log Z,\eqno(1)$$
\noindent
with a r.m.s. of 0.05 mag, whereas fr $Z<$ 0.008 we get
$$c=-0.23(\pm0.03)\log (Z/0.008)\eqno(2)$$
\noindent
Note that these corrections hold only in the framework
of the KP method, i.e. they apply to distance moduli inferred from
LMC-based $V$ and $I$ linear PL relations, according to the
procedure discussed above.

Figure 4 depicts the dependence of the predicted metallicity
corrections on $Z$, for selected reasonable $\Delta Y$/$\Delta Z$
ratios. In order to test these corrections,  which are inferred by
fundamental pulsation models, we take into consideration the HST
galaxies which give Cepheid-calibrated SNIa luminosities and we
evaluate the Cepheid true distance modulus $\mu^L_{0,Z,Y}$ using the
predicted $V$ and $I$ PL relations at the various chemical
compositions. Figure 5 shows that the average differences
$<\mu^L_{0,Z,Y}-\mu^L_{0,0.008}>$  
among the distance moduli provided
by linear relations (see data listed in Table 7) 
are in close agreement with the predicted corrections given in eq. (1)
and eq. (2). 

The results plotted in Fig. 4 show several points worthy of
notice: {\it i}) - the predicted correction is not linear with
log$Z$ but shows a ``turn-over'' around the solar metal content,
depending on the helium-to-hydrogen enrichment ratio.
Specifically, the ``turn-over'' metallicity $Z_{to}$ increases
from $\sim$ 0.016 to 0.025 as $\Delta Y$/$\Delta Z$ decreases from
3.5 to 2.0; {\it ii}) - if LMC-calibrated PL relations are used,
then Cepheids with 0.008$\le Z\le Z_{to}$ result spuriously
brighter and, consequently, the LMC-based distance modulus may
require a ${\it negative}$ correction whose amount increases for
larger $Z$ (at fixed $\Delta Y/\Delta Z$) and lower
helium-to-hydrogen ratio (at fixed $Z$); {\it iii}) - this
behavior is reversed when the Cepheid metallicity is larger than
$Z_{to}$, and eventually the LMC-based distance modulus of
metal-rich Cepheids may need a ${\it positive}$ correction whose
amount increases for larger $Z$ (at fixed $\Delta Y/\Delta Z$) and
higher helium-to-hydrogen ratio (at fixed $Z$); {\it iv}) -
depending on $\Delta Y$/$\Delta Z$, very metal-rich ($Z\sim$ 0.04)
Cepheids may appear spuriously brighter or fainter. We estimate
that the metallicity correction to the LMC-based distance modulus
of these variables varies from $\sim$+0.25 to $\sim-$0.15 mag as
the helium-to-metal enrichment ratio decreases from 3.5 to 2.0,
respectively.

Taking advantage of present results, we can try to solve the discrepancy
between the empirical metallicity correction based on M101
observations and the theoretical one given by Caputo et al.
(2000b). Kennicutt et al. (1998) have
used the Wide Field Planetary Camera 2 (WFPC2) on HST to observe
Cepheids in two fields of this galaxy that span a range in oxygen
abundance of $\sim$ 0.7 dex, with the outer field showing a
LMC-like value. According to Kennicutt et al.
(1998), the true distance modulus inferred from LMC-based
$V$ and $I$ PL relations is  29.21$\pm$0.09
mag and 29.36$\pm$0.08 mag for the inner and the outer field,
respectively, leading to a difference of 0.16$\pm$0.10 mag over an
abundance baseline of $\Delta$[O/H]=0.68$\pm$0.15 dex,
and to the already mentioned correction for metallicity of
$\sim$0.24($\pm$0.16)$\Delta$[O/H].

Fig. 6 shows the predicted corrections in Fig. 4 as a function of
[O/H], assuming [O/H]=log$(Z/X)-$log$(Z/X)_{\odot}$, and with
$Z_{\odot}$=0.02 and $Y_{\odot}$=0.27 (see data in Table 8).
The two vertical arrows depict the average oxygen abundance of the
outer and inner fields in M101, while the horizontal arrow refers
to the difference of 0.16 mag between the LMC-referenced distance
moduli. At the light of present theoretical results, we derive
that the predicted metallicity correction to the inner field
distance is negative with $\Delta Y$/$\Delta Z\le$ 2.5 {\it but}
positive with $\Delta Y$/$\Delta Z\ge$ 3.0, reaching $\sim$ 0.20
mag with [O/H]$\sim$0.35 and $\Delta Y$/$\Delta Z$=3.5.

\section{Conclusions}

Pulsating nonlinear convective models spanning wide ranges of helium and
metal content are used to study the behavior of the Cepheid instability
strip and the dependence of the
Cepheid distance scale on the chemical composition of the pulsators.
We show that the combined effects of
helium and metal abundances have to be simultaneously considered
in order to properly address the correction to
Cepheid
distances inferred from $V$ and $I$ LMC-referenced PL relations,
mainly for Cepheids at solar and sovra-solar metal
abundances, where the helium content is significantly
dependent on the
adopted helium-to-hydrogen ratio $\Delta Y/\Delta Z$.

The main results may be listed as follows:
\begin{enumerate}
\item   The adoption
of {\it universal} LMC-based PL$_V$ and PL$_I$ linear relations
 to get distance moduli within $\pm$ 0.10 mag is fully
justified for Cepheids in the short-period range ($P <$ 10 d),
almost independent of chemical composition.

\item   For Cepheids in the long period range (20-30 days),
the adoption of the LMC-referenced linear relations may provide
the distance modulus with an uncertainty of $\pm$0.1 mag, on
condition that the helium-to-hydrogen enrichment ratio is
reasonably high ($\Delta Y/\Delta Z\ge$ 3).

\item    As a whole, for Cepheids with $P>$ 10 days and $Z>0.008$ the estimated
correction (in mag) to the intrinsic distance modulus inferred
from LMC-based PL$_V$ and PL$_I$ linear relations can be
approximated as $c=-6.03+17.80Y-2.44Y-2.80$log$Z+7.33Y$log$Z$,
with an intrinsic uncertainty of 0.05 mag. For Cepheids with $Z<$
0.008, the correction is $c=-0.23(\pm$0.03)log$(Z/0.008)$.

\item   Taking
advantage of present results, we give evidence that the
sensitivity of the Cepheid predicted PL$_V$ and PL$_I$ relations
to the helium content could account for the Kennicutt et al.
(1998) empirical metallicity correction based on M101
observations, on condition that  the helium-to-hydrogen
enrichment ratio is $\Delta Y/\Delta Z \sim$ 3.5.

\item   Even though
discussion on the actual value of the ratio $\Delta Y/\Delta Z$ is
beyond
the purpose of this paper,  and the uncertainty
on current estimates is still disappointingly large,
such a value is consistent with
a ratio  of 4$\pm$1 (Pagel et al. 1992) and of 3$\pm$2 (Pagel \&
Portinari 1998) determined from extragalactic HII regions and
nearby main-sequence stars, respectively. On the other hand,
a ratio $\Delta Y/\Delta Z\sim$ 2 has been suggested by
Izotov, Thuan, \& Lipovetsky (1997).

\item   In summary,
reliable estimates of the ratio $\Delta Y/\Delta Z$ are needed to
properly evaluate the effects of $Y$ and $Z$ on the LMC-based
distance modulus of Cepheids with $P>$ 10 days. However, according
to the data plotted in Fig. 6, we can conclude that over the range
in [O/H] covered by HST galaxies ($-$0.75 to +0.35) the hypothesis
of a constant helium-to-hydrogen ratio seems to exclude that the
correction to the LMC-based distance moduli is within $\pm$0.1 mag
for all galaxies, whichever is the value of the $\Delta Y/\Delta
Z$ ratio.
\end{enumerate}

\acknowledgements

We deeply thank our referee for his/her several comments and
suggestions. This work was supported by MIUR-Cofin 2000, under the
scientific project "Stellar Observables of Cosmological
Relevance".

\pagebreak

\clearpage

\begin{table}
\begin{center}

\caption{Intrinsic parameters and intensity averaged
mean magnitudes and colors for models at  $Z$=0.02 and $Y$=0.31}
\vspace{0.3 cm}

\begin{tabular}{cclccccccc}
\hline
\hline
M/M$\odot$ & T$_{eff}$ &L/L$\odot$& P     & V & B$-$V & V$-$K & V$-$I & V$-$J & V$-$R \\
           &    [K]    &          & [day] &  mag & mag& mag & mag& mag & mag\\
\hline
  5 &5900 & 3.13 & 3.7352 &  $-$3.0659  &  0.6015  &  1.3995  &  0.6539 &  1.0472  &  0.3294  \\
  5 &5800 & 3.13 & 3.9523 &  $-$3.0494  &  0.6403  &  1.4692  &  0.6858 &  1.0988  &  0.3465  \\
  5 &5700 & 3.13 & 4.1741 &  $-$3.0322  &  0.6795  &  1.5370  &  0.7167 & 1.1493  &  0.3633  \\
  5 &5600 & 3.13 & 4.4386 &  $-$3.0118  &  0.7241  &  1.6121  &  0.7508 & 1.2051  &  0.3821  \\
  5 &5500 & 3.13 & 4.7083 &  $-$2.9908  &  0.7669  &  1.6851  &  0.7831 & 1.2595  &  0.3999  \\
  5 &5400 & 3.13 & 4.9924 &  $-$2.9683  &  0.8096  &  1.7586  &  0.8151 &  1.3145  &  0.4175  \\
  7 &5600 & 3.62 & 9.3579 &  $-$4.2432  &  0.7306  &  1.6032  &  0.7495 &  1.2015  &  0.3816  \\
  7 &5400 & 3.62 & 10.676 &  $-$4.1863  &  0.8000  &  1.7834  &  0.8139 &  1.3283  &  0.4164  \\
  7 &5300 & 3.62 & 11.293 &  $-$4.1565  &  0.8372  &  1.8715  &  0.8464 &  1.3909  &  0.4340  \\
  7 &5200 & 3.62 & 12.063 &  $-$4.1214  &  0.8806  &  1.9684  &  0.8828 &  1.4600  &  0.4539  \\
  7 &5100 & 3.62 & 12.870 &  $-$4.0857  &  0.9269  &  2.0622  &  0.9195 &  1.5281  &  0.4743  \\
  7 &5000 & 3.62 & 13.745 &  $-$4.0487  &  0.9761  &  2.1541  &  0.9565 &  1.5959  &  0.4951  \\
  7 &4900 & 3.62 & 14.645 &  $-$4.0146  &  1.0301  &  2.2345  &  0.9921 &  1.6580  &  0.5156  \\
  9 &5300 & 3.98 & 19.972 &  $-$5.0584  &  0.8427  &  1.8677  &  0.8469 &  1.3915  &  0.4349  \\
  9 &5200 & 3.98 & 21.310 &  $-$5.0235  &  0.8872  &  1.9633  &  0.8832 &  1.4605  &  0.4550  \\
  9 &5100 & 3.98 & 22.788 &  $-$4.9884  &  0.9409  &  2.0557  &  0.9218 &  1.5292  &  0.4767  \\
  9 &5000 & 3.98 & 24.361 &  $-$4.9511  &  0.9969  &  2.1474  &  0.9606 &  1.5981  &  0.4988  \\
  9 &4900 & 3.98 & 26.023 &  $-$4.9113  &  1.0529  &  2.2410  &  0.9996 &  1.6687  &  0.5210  \\
  9 &4800 & 3.98 & 27.876 &  $-$4.8670  &  1.1069  &  2.3406  &  1.0392 &  1.7427  &  0.5434  \\
  9 &4700 & 3.98 & 29.833 &  $-$4.8203  &  1.1593  &  2.4408  &  1.0784 &  1.8168  &  0.5653  \\
  9 &4600 & 3.98 & 31.992 &  $-$4.7725  &  1.2131  &  2.5389  &  1.1176 &  1.8904  &  0.5875  \\
  9 &4500 & 3.98 & 34.336 &  $-$4.7250  &  1.2673  &  2.6318  &  1.1552 &  1.9616  &  0.6088  \\
 11 &5100 & 4.27 & 36.206 &  $-$5.7246  &  0.9910  &  2.0269  &  0.9339 &  1.5235  &  0.4862  \\
 11 &5000 & 4.27 & 38.948 &  $-$5.6769  &  1.0167  &  2.1449  &  0.9662 &  1.6023  &  0.5031  \\
 11 &4900 & 4.27 & 41.728 &  $-$5.6323  &  1.0645  &  2.2494  &  1.0046 &  1.6779  &  0.5243  \\
 11 &4800 & 4.27 & 44.745 &  $-$5.5856  &  1.1170  &  2.3544  &  1.0450 &  1.7547  &  0.5468  \\
 11 &4700 & 4.27 & 47.909 &  $-$5.5355  &  1.1709  &  2.4610  &  1.0866 &  1.8329  &  0.5699  \\
 11 &4600 & 4.27 & 51.469 &  $-$5.4824  &  1.2249  &  2.5690  &  1.1289 &  1.9121  &  0.5933  \\
 11 &4500 & 4.27 & 55.241 &  $-$5.4268  &  1.2779  &  2.6783  &  1.1719 &  1.9922  &  0.6171  \\
 11 &4400 & 4.27 & 59.377 &  $-$5.3677  &  1.3297  &  2.7896  &  1.2161 &  2.0735  &  0.6411  \\
 11 &4300 & 4.27 & 63.752 &  $-$5.3084  &  1.3777  &  2.8973  &  1.2597 &  2.1522  &  0.6645  \\
 11 &4200 & 4.27 & 68.402 &  $-$5.2498  &  1.4229  &  3.0011  &  1.3030 &  2.2280  &  0.6878  \\
\hline
\end{tabular}

\end{center}

\end{table}

\begin{table}
\begin{center}

\caption{The same as Table 1 but with $Z$=0.03 and $Y$=0.31}
\vspace{0.3 cm}
\begin{tabular}{cclccccccc}
\hline
\hline
M/M$\odot$ & T$_{eff}$ &L/L$\odot$& P     & V & B-V & V-K & V-I & V-J & V-R \\
           &    [K]    &          & [day] &  mag & mag& mag & mag& mag & mag\\
\hline
5  &5800  & 3.07  & 3.5149   &  $-$2.9056  &  0.6703  &   1.4608  &  0.7060  &  1.1010  &  0.3734 \\
5  &5600  & 3.07  & 3.9492   &  $-$2.8640  &  0.7368  &     1.6103  &  0.7676  &  1.2084  &  0.4066 \\
5  &5500  & 3.07  & 4.1926   &  $-$2.8412  &  0.7804  &     1.6872  &  0.8020  &  1.2657  &  0.4260 \\
5  &5400  & 3.07  & 4.4544   &  $-$2.8162  &  0.8259  &     1.7661  &  0.8374  &  1.3249  &  0.4459 \\
5  &5300  & 3.07  & 4.7357   &  $-$2.7893  &  0.8718  &     1.8461  &  0.8728  &  1.3851  &  0.4657 \\
5  &5200  & 3.07  & 5.0348   &  $-$2.7601  &  0.9179  &     1.9278  &  0.9080  &  1.4468  &  0.4853 \\
7  &5300  & 3.56  & 1.0063   &  $-$4.0151  &  0.8757  &     1.8487  &  0.8726  &  1.3882  &  0.4656 \\
7  &5200  & 3.56  & 1.0738   &  $-$3.9791  &  0.9089  &     1.9473  &  0.9070  &  1.4569  &  0.4837 \\
7  &5100  & 3.56  & 1.1495   &  $-$3.9391  &  0.9478  &     2.0516  &  0.9450  &  1.5298  &  0.5041 \\
7  &5000  & 3.56  & 1.2245   &  $-$3.8983  &  0.9896  &     2.1535  &  0.9837  &  1.6017  &  0.5252 \\
7  &4900  & 3.56  & 1.3145   &  $-$3.8514  &  1.0414  &     2.2641  &  1.0282  &  1.6818  &  0.5497 \\
7  &4800  & 3.56  & 1.4093   &  $-$3.8050  &  1.0938  &     2.3665  &  1.0711  &  1.7574  &  0.5736 \\
7  &4700  & 3.56  & 1.5064   &  $-$3.7610  &  1.1526  &     2.4594  &  1.1132  &  1.8293  &  0.5977 \\
7  &4600  & 3.56  & 1.6068   &  $-$3.7256  &  1.2238  &     2.5273  &  1.1500  &  1.8911  &  0.6199 \\
9  &4800  & 3.92  & 2.4982   &   $-$4.7085 &   1.1260 &     2.3554  &  1.0759  &  1.7611  &  0.5785 \\
9  &4700  & 3.92  & 2.6803   &   $-$4.6562 &   1.1802 &     2.4660  &  1.1211  &  1.8421  &  0.6034 \\
9  &4600  & 3.92  & 2.8851   &   $-$4.5993 &   1.2335 &     2.5820  &  1.1682  &  1.9263  &  0.6291 \\
9  &4500  & 3.92  & 3.0662   &   $-$4.5431 &   1.2772 &     2.6929  &  1.2126  &  2.0048  &  0.6527 \\
9  &4400  & 3.92  & 3.3295   &   $-$4.4773 &   1.3400 &     2.8143  &  1.2657  &  2.0964  &  0.6820 \\
9  &4300  & 3.92  & 3.5804   &   $-$4.4167 &   1.3930 &     2.9219  &  1.3139  &  2.1775  &  0.7086 \\
11 & 4700 &  4.21 &  4.3504  &  $-$5.3932  &  1.2083  &     2.4639  &  1.1270  &  1.8500  &  0.6083 \\
11 & 4600 &  4.21 &  4.6861  &  $-$5.3308  &  1.2496  &     2.5926  &  1.1748  &  1.9384  &  0.6330 \\
11 & 4500 &  4.21 &  5.0461  &  $-$5.2663  &  1.3014  &     2.7176  &  1.2252  &  2.0277  &  0.6599 \\
11 & 4400 &  4.21 &  5.4382  &  $-$5.1969  &  1.3555  &     2.8471  &  1.2786  &  2.1209  &  0.6885 \\
11 & 4300 &  4.21 &  5.8492  &  $-$5.1263  &  1.4070  &     2.9731  &  1.3317  &  2.2119  &  0.7165 \\
11 & 4200 &  4.21 &  6.2925  &  $-$5.0528  &  1.4572  &     3.0999  &  1.3865  &  2.3037  &  0.7453 \\
11 & 4100 &  4.21 &  6.7782  &  $-$4.9783  &  1.5045  &     3.2247  &  1.4418  &  2.3940  &  0.7743 \\
11 & 4000 &  4.21 &  7.3029  &  $-$4.9012  &  1.5506  &     3.3496  &  1.4989  &  2.4849  &  0.8044 \\
\hline
\end{tabular}
\end{center}
\end{table}

\begin{table}
\begin{center}

\caption{The same as Table 1 but with $Z$=0.04 and $Y$=0.33}
\vspace{0.3 cm}
\begin{tabular}{cclccccccc}
\hline
\hline
M/M$\odot$ & T$_{eff}$ &L/L$\odot$& P     & V & B-V & V-K & V-I & V-J & V-R \\
           &    [K]    &          & [day] &  mag & mag& mag & mag& mag & mag\\
\hline
 5  &5600  & 3.06  & 3.9097  & $-$2.8446   & 0.7689 &   1.6022 &   0.7903 &   1.2112  &  0.4355  \\
 5  &5500  & 3.06  & 4.1402  & $-$2.8214   & 0.8088 &  1.6781  &  0.8244  &  1.2682   & 0.4548   \\
 5  &5400  & 3.06  & 4.4214  & $-$2.7955   & 0.8515 &  1.7575  &  0.8602  &  1.3281   & 0.4749   \\
 5  &5300  & 3.06  & 4.6926  & $-$2.7676   & 0.8963 &  1.8377  &  0.8966  &  1.3889   & 0.4953   \\
 7  &5100  & 3.55  & 1.1410  & $-$3.9160   & 0.9484 &  2.0548  &  0.9668  &  1.5359   & 0.5303   \\
 7  &5000  & 3.55  & 1.2252  & $-$3.8732   & 1.0068 &  2.1596  &  1.0136  &  1.6143   & 0.5568   \\
 7  &4900  & 3.55  & 1.3096  & $-$3.8374   & 1.0753 &  2.2407  &  1.0569  &  1.6813   & 0.5828   \\
 7  &4800  & 3.55  & 1.3934  & $-$3.8012   & 1.1444 &  2.3195  &  1.0985  &  1.7482   & 0.6077   \\
 9  &4800  & 3.92  & 2.4912  & $-$4.7145   & 1.1599 &  2.3169  &  1.1015  &  1.7517   & 0.6100   \\
 9  &4700  & 3.92  & 2.7091  & $-$4.6521   & 1.2142 &  2.4497  &  1.1548  &  1.8471   & 0.6382   \\
 9  &4600  & 3.92  & 2.9120  & $-$4.5940   & 1.2654 &  2.5675  &  1.2039  &  1.9327   & 0.6645   \\
 9  &4500  & 3.92  & 3.1200  & $-$4.5343   & 1.3171 &  2.6831  &  1.2536  &  2.0175   & 0.6913   \\
 9  &4400  & 3.92  & 3.3660  & $-$4.4727   & 1.3690 &  2.7977  &  1.3045  &  2.1023   & 0.7188   \\
 11 & 4600 &  4.21 &  4.7210 & $-$5.3205   & 1.2868 &   2.5713 &   1.2100 &   1.9423  &  0.6686  \\
 11 & 4500 &  4.21 &  5.0738 & $-$5.2541   & 1.3321 &   2.7016 &   1.2628 &   2.0336  &  0.6964  \\
 11 & 4400 &  4.21 &  5.4642 & $-$5.1848   & 1.3822 &   2.8306 &   1.3178 &   2.1261  &  0.7255  \\
 11 & 4300 &  4.21 &  5.8821 & $-$5.1130   & 1.4322 &   2.9590 &   1.3741 &   2.2188  &  0.7552  \\
 11 & 4200 &  4.21 &  6.3580 & $-$5.0370   & 1.4822 &   3.0903 &   1.4332 &   2.3139  &  0.7865  \\
 11 & 4100 &  4.21 &  6.8477 & $-$4.9614   & 1.5287 &   3.2160 &   1.4914 &   2.4053  &  0.8177  \\
\hline
\end{tabular}
\end{center}
\end{table}

\begin{table}
\caption{Theoretical PLC relations for fundamental pulsators.}
\begin{center}
\begin{tabular}{cccccc}
\hline
\hline
Z & Y & $\alpha$ & $\beta$  & $\gamma$ & $\sigma$\\
\hline\\
  \multicolumn{5}{c}{$<M_V>$=$\alpha$+$\beta$log$P$+$\gamma$[$<B>-<V>$]} \\
0.004 & 0.25 &$-$2.54 $\pm$0.04& $-$3.52 $\pm$0.03& 2.79 $\pm$0.07&  0.04\\
0.008 & 0.25 &$-$2.63 $\pm$0.04& $-$3.55 $\pm$0.03& 2.83 $\pm$0.06& 0.03\\
0.02  & 0.28 &$-$2.98 $\pm$0.07& $-$3.72 $\pm$0.10& 3.27 $\pm$0.18& 0.07\\
0.03  & 0.31 &$-$3.10 $\pm$0.06& $-$3.81 $\pm$0.08& 3.34 $\pm$0.13& 0.06\\
0.04  & 0.33 &$-$3.12 $\pm$0.14& $-$3.76 $\pm$0.15& 3.24 $\pm$0.28& 0.14\\
0.02  & 0.31 &$-$2.79 $\pm$0.03& $-$3.79 $\pm$0.04& 3.10 $\pm$0.07&  0.03\\
    \multicolumn{5}{c}{$<M_V>$=$\alpha$+$\beta$log$P$+$\gamma$[$<V>-<R>$]} \\
0.004 & 0.25 &$-$3.28 $\pm$0.03& $-$3.57 $\pm$0.02& 6.93 $\pm$0.12& 0.03 \\
0.008 & 0.25 &$-$3.31 $\pm$0.03& $-$3.59 $\pm$0.02& 6.97 $\pm$0.11& 0.03 \\
0.02  & 0.28 &$-$3.40 $\pm$0.04& $-$3.62 $\pm$0.05& 7.09 $\pm$0.18& 0.04 \\
0.03  & 0.31 &$-$3.39 $\pm$0.02& $-$3.69 $\pm$0.02& 6.66 $\pm$0.06& 0.02\\
0.04  & 0.33 &$-$3.44 $\pm$0.04& $-$3.64 $\pm$0.03& 6.31 $\pm$0.12& 0.04\\
0.02  & 0.31 &$-$3.31 $\pm$0.02& $-$3.74 $\pm$0.02& 7.13 $\pm$0.08&  0.02\\
    \multicolumn{5}{c}{$<M_V>$=$\alpha$+$\beta$log$P$+$\gamma$[$<V>-<I>$]} \\
0.004 & 0.25 &$-$3.55 $\pm$0.03& $-$3.58 $\pm$0.03& 3.75 $\pm$0.07& 0.03 \\
0.008 & 0.25 &$-$3.54 $\pm$0.03& $-$3.59 $\pm$0.02& 3.74 $\pm$0.06& 0.03 \\
0.02  & 0.28 &$-$3.61 $\pm$0.03& $-$3.59 $\pm$0.04& 3.85 $\pm$0.09& 0.03 \\
0.03  & 0.31 &$-$3.45 $\pm$0.01& $-$3.65 $\pm$0.01& 3.58 $\pm$0.03& 0.01\\
0.04  & 0.33 &$-$3.38 $\pm$0.02& $-$3.61 $\pm$0.01& 3.39 $\pm$0.03& 0.02\\
0.02  & 0.31 &$-$3.56 $\pm$0.02& -3.73 $\pm$0.02& 3.94 $\pm$0.04&  0.02\\
    \multicolumn{5}{c}{$<M_V>$=$\alpha$+$\beta$log$P$+$\gamma$[$<V>-<J>$]} \\
0.004 & 0.25 &$-$3.47 $\pm$0.03& $-$3.60 $\pm$0.03& 2.26 $\pm$0.04& 0.03 \\
0.008 & 0.25 &$-$3.35 $\pm$0.05& $-$3.60 $\pm$0.05& 2.17 $\pm$0.07& 0.05 \\
0.02  & 0.28 &$-$3.29 $\pm$0.04& $-$3.59 $\pm$0.04& 2.11 $\pm$0.05& 0.03 \\
0.03  & 0.31 &$-$3.18 $\pm$0.02& $-$3.71 $\pm$0.02& 2.08 $\pm$0.02& 0.02\\
0.04  & 0.33 &$-$3.18 $\pm$0.03& $-$3.70 $\pm$0.03& 2.08 $\pm$0.03& 0.03\\
0.02  & 0.31 &$-$3.17 $\pm$0.01& $-$3.72 $\pm$0.01& 2.12 $\pm$0.02&  0.01\\
    \multicolumn{5}{c}{$<M_V>$=$\alpha$+$\beta$log$P$+$\gamma$[$<V>-<K>$]} \\
0.004 & 0.25 &$-$3.44 $\pm$0.04& $-$3.61 $\pm$0.03& 1.64 $\pm$0.03& 0.04\\
0.008 & 0.25 &$-$3.37 $\pm$0.04& $-$3.60 $\pm$0.03& 1.61 $\pm$0.03& 0.03 \\
0.02  & 0.28 &$-$3.25 $\pm$0.04& $-$3.55 $\pm$0.05& 1.53 $\pm$0.04& 0.04\\
0.03  & 0.31 &$-$3.10 $\pm$0.02& $-$3.67 $\pm$0.03& 1.50 $\pm$0.02& 0.02\\
0.04  & 0.33 &$-$3.10 $\pm$0.04& $-$3.65 $\pm$0.04& 1.50 $\pm$0.04& 0.04\\
0.02  & 0.31 &$-$3.13 $\pm$0.02& $-$3.69 $\pm$0.02& 1.54 $\pm$0.02&  0.02\\
\hline
\end{tabular}
\end{center}
\end{table}

\tablewidth{0pt}
\begin{deluxetable}{cccccc}
\tablecaption{Theoretical PL relations for fundamental pulsators.
Quadratic solutions: $\overline{M_{\lambda}}$=$a$+$b$log$P$+$c$log$P^2$.}
\tablehead{
\colhead{Z}&
\colhead{Y}&
\colhead{$a$}&
\colhead{$b$}&
\colhead{$c$}&
\colhead{$\sigma$}}
\startdata
\multicolumn{5}{c}{$\overline{M_B}$}\nl
0.004 & 0.25 &$-$0.01$\pm$0.06&$-$4.81$\pm$0.10&1.14$\pm$0.06 & 0.24\nl
0.008 & 0.25 &$-$0.21$\pm$0.06&$-$4.17$\pm$0.13&0.94$\pm$0.06 & 0.26\nl
0.02 & 0.28 &$-$0.93$\pm$0.04&$-$2.43$\pm$0.09&0.39$\pm$0.04  & 0.21\nl
0.03 & 0.31 &$-$0.88$\pm$0.06&$-$2.05$\pm$0.12&0.27$\pm$0.05  & 0.24\nl
0.04 & 0.33 &$-$0.59$\pm$0.03&$-$2.40$\pm$0.07&0.36$\pm$0.03  & 0.13\nl
0.02 & 0.31 &$-$0.66$\pm$0.06&$-$3.00$\pm$0.10&0.54$\pm$0.06  & 0.24\nl
\multicolumn{5}{c}{$\overline{M_V}$}\nl
0.004 & 0.25 &$-$0.69$\pm$0.04&$-$4.43$\pm$0.10&0.81$\pm$0.05 & 0.18\nl
0.008 & 0.25 &$-$0.86$\pm$0.04&$-$3.98$\pm$0.09&0.67$\pm$0.05 & 0.19\nl
0.02 & 0.28 &$-$1.41$\pm$0.03&$-$2.75$\pm$0.07&0.30$\pm$0.03  & 0.16\nl
0.03 & 0.31 &$-$1.36$\pm$0.04&$-$2.54$\pm$0.09&0.24$\pm$0.04  & 0.18 \nl
0.04 & 0.33 &$-$1.16$\pm$0.02&$-$2.78$\pm$0.05&0.30$\pm$0.02  & 0.10 \nl
0.02 & 0.31 & $-$1.18$\pm$0.05&$-$3.20$\pm$0.10&0.40$\pm$0.05 & 0.18 \nl
\multicolumn{5}{c}{$\overline{M_R}$}\nl
0.004 & 0.25 &$-$1.08$\pm$0.04&$-$4.28$\pm$0.08&0.67$\pm$0.04 & 0.15  \nl
0.008 & 0.25 &$-$1.22$\pm$0.04&$-$3.91$\pm$0.08&0.56$\pm$0.04 & 0.16 \nl
0.02 & 0.28 &$-$1.69$\pm$0.03&$-$2.88$\pm$0.06&0.26$\pm$0.03  & 0.13 \nl
0.03 & 0.31 &$-$1.67$\pm$0.04&$-$2.71$\pm$0.07&0.20$\pm$0.03  & 0.15  \nl
0.04 & 0.33 &$-$1.53$\pm$0.02&$-$2.92$\pm$0.04&0.25$\pm$0.02  & 0.08 \nl
0.02 & 0.31 &$-$1.48$\pm$0.04&$-$3.28$\pm$0.08&0.34$\pm$0.04  & 0.15 \nl
\multicolumn{5}{c}{$\overline{M_I}$}\nl
0.004 & 0.24 &$-$1.48$\pm$0.03&$-$4.16$\pm$0.07&0.56$\pm$0.03 & 0.13 \nl
0.008 & 0.24 &$-$1.59$\pm$0.03&$-$3.84$\pm$0.07&0.47$\pm$0.03 & 0.13 \nl
0.02 & 0.28 &$-$1.98$\pm$0.02&$-$2.99$\pm$0.05&0.23$\pm$0.02  & 0.11 \nl
0.03 & 0.31 &$-$1.96$\pm$0.03&$-$2.83$\pm$0.06&0.16$\pm$0.03  & 0.13  \nl
0.04 & 0.33 &$-$1.84$\pm$0.02&$-$3.00$\pm$0.03&0.20$\pm$0.02  & 0.07 \nl
0.02 & 0.31 &$-$1.78$\pm$0.03&$-$3.35$\pm$0.07&0.30$\pm$0.03  & 0.13 \nl
\multicolumn{5}{c}{$\overline{M_J}$}\nl
0.004 & 0.24 &$-$1.97$\pm$0.02&$-$3.97$\pm$0.05&0.40$\pm$0.02  & 0.09  \nl
0.008 & 0.24 &$-$2.04$\pm$0.02&$-$3.73$\pm$0.05&0.33$\pm$0.02  & 0.10 \nl
0.02 & 0.28 &$-$2.32$\pm$0.01&$-$3.08$\pm$0.03&0.12$\pm$0.02   & 0.08 \nl
0.03 & 0.31 &$-$2.27$\pm$0.02&$-$3.02$\pm$0.04&0.09$\pm$0.02   & 0.09 \nl
0.04 & 0.33 &$-$2.17$\pm$0.01&$-$3.15$\pm$0.02&0.12$\pm$0.01   & 0.05 \nl
0.02 & 0.31 &$-$2.15$\pm$0.02&$-$3.37$\pm$0.05&0.17$\pm$0.02   & 0.09 \nl
\multicolumn{5}{c}{$\overline{M_K}$}\nl
0.004 & 0.24 &$-$2.42$\pm$0.01&$-$3.81$\pm$0.03&0.26$\pm$0.01 & 0.06\nl
0.008 & 0.24 &$-$2.45$\pm$0.01&$-$3.65$\pm$0.03&0.21$\pm$0.01 & 0.06\nl
0.02 & 0.28 &$-$2.63$\pm$0.01&$-$3.19$\pm$0.02&0.06$\pm$0.01  & 0.05\nl
0.03 & 0.31 &$-$2.59$\pm$0.01&$-$3.15$\pm$0.03&0.01$\pm$0.01  & 0.06 \nl
0.04 & 0.33 &$-$2.51$\pm$0.01&$-$3.23$\pm$0.01&0.04$\pm$0.01  & 0.03 \nl
0.02 & 0.31 &$-$2.48$\pm$0.01&$-$3.43$\pm$0.03&0.10$\pm$0.01  & 0.06 \nl
&&&&\\
\enddata
\end{deluxetable}

\tablewidth{0pt}
\begin{deluxetable}{ccccc}
\tablecaption{Theoretical PL relations for fundamental pulsators.
Linear solutions with log$P\le$ 1.5:
$\overline{M_{\lambda}}$=$a$+$b$log$P$. } \tablehead{ \colhead{Z}&
\colhead{Y}& \colhead{$a$}& \colhead{$b$}& \colhead{$\sigma$}}
\startdata \multicolumn{5}{c}{$\overline{M_B}$}\nl
0.004& 0.25 &$-$0.90$\pm$0.03&$-$2.71$\pm$0.02&0.24\nl
0.008& 0.25 &$-$0.93$\pm$0.03&$-$2.44$\pm$0.02&0.25\nl
0.02& 0.28&$-$1.21$\pm$0.02&$-$1.73$\pm$0.02&0.19\nl
0.03 & 0.31&$-$1.08$\pm$0.02&$-$1.55$\pm$0.02&0.23\nl
0.04 & 0.33&$-$0.85$\pm$0.01&$-$1.76$\pm$0.01&0.12\nl
0.02 & 0.31&$-$1.08$\pm$0.02&$-$2.00$\pm$0.02&0.23\nl
\multicolumn{5}{c}{$\overline{M_V}$}\nl
0.004 & 0.25&$-$1.32$\pm$0.02&$-$2.94$\pm$0.02&0.17\nl
0.008 & 0.25&$-$1.37$\pm$0.02&$-$2.75$\pm$0.02&0.18\nl
0.02 & 0.28&$-$1.62$\pm$0.01&$-$2.22$\pm$0.01&0.14\nl
0.03 & 0.31&$-$1.54$\pm$0.01&$-$2.10$\pm$0.01&0.17\nl
0.04 & 0.33&$-$1.38$\pm$0.01&$-$2.26$\pm$0.01&0.09\nl
0.02 & 0.31&$-$1.50$\pm$0.02&$-$2.45$\pm$0.02&0.17\nl
\multicolumn{5}{c}{$\overline{M_R}$}\nl
0.004 & 0.25 &$-$1.61$\pm$0.02&$-$3.03$\pm$0.01&0.15\nl
0.008 & 0.25 &$-$1.65$\pm$0.02&$-$2.87$\pm$0.01&0.16\nl
0.02 & 0.28 &$-$1.88$\pm$0.01&$-$2.42$\pm$0.01&0.12\nl
0.03 & 0.31&$-$1.82$\pm$0.01&$-$2.34$\pm$0.01&0.15\nl
0.04 & 0.33&$-$1.71$\pm$0.01&$-$2.48$\pm$0.01&0.08\nl
0.02 & 0.31&$-$1.75$\pm$0.01&$-$2.64$\pm$0.02&0.15\nl
\multicolumn{5}{c}{$\overline{M_I}$}\nl
0.004 & 0.25&$-$1.92$\pm$0.01&$-$3.11$\pm$0.01&0.12\nl
0.008 & 0.25&$-$1.95$\pm$0.01&$-$2.98$\pm$0.01&0.13\nl
0.02 & 0.28&$-$2.14$\pm$0.01&$-$2.58$\pm$0.01&0.10\nl
0.03 & 0.31&$-$2.08$\pm$0.01&$-$2.53$\pm$0.01&0.12\nl
0.04 & 0.33&$-$1.98$\pm$0.01&$-$2.65$\pm$0.01&0.06\nl
0.02 & 0.31&$-$2.02$\pm$0.01&$-$2.78$\pm$0.01&0.13\nl
\multicolumn{5}{c}{$\overline{M_J}$}\nl 0.004 & 0.25
&$-$2.28$\pm$0.01&$-$3.23$\pm$0.01&0.09\nl 0.008 & 0.25 &
$-$2.29$\pm$0.01&$-$3.13$\pm$0.01&0.10\nl 0.02 & 0.28 &
$-$2.41$\pm$0.01&$-$2.87$\pm$0.01&0.07\nl 0.03 & 0.31
&$-$2.34$\pm$0.01&$-$2.87$\pm$0.01&0.09\nl 0.04 & 0.33
&$-$2.26$\pm$0.01&$-$2.94$\pm$0.01&0.04\nl 0.02 & 0.31
&$-$2.29$\pm$0.01&$-$3.05$\pm$0.01&0.09\nl
\multicolumn{5}{c}{$\overline{M_K}$}\nl 0.004 & 0.25 &
$-$2.61$\pm$0.01&$-$3.33$\pm$0.01&0.06\nl 0.008 & 0.25 &
$-$2.61$\pm$0.01&$-$3.27$\pm$0.01&0.06\nl 0.02 & 0.28 &
$-$2.67$\pm$0.01&$-$3.09$\pm$0.01&0.04\nl 0.03 & 0.31
&$-$2.60$\pm$0.01&$-$3.12$\pm$0.01&0.05\nl 0.04 & 0.33
&$-$2.54$\pm$0.01&$-$3.16$\pm$0.01&0.03\nl 0.02 & 0.31
&$-$2.55$\pm$0.01&$-$3.24$\pm$0.01&0.06\nl
\enddata
\end{deluxetable}

\begin{table}
\begin{center}
\caption{Difference between $\mu^L_{0,Z,Y}$ and $\mu^L_{0,0.008}$
for SNIa host galaxies.}

\vspace{0.5truecm}

\begin{tabular}{lccccc}
\hline
\hline
galaxy & Z=0.004 &  Z=0.02 &  Z=0.03 & Z=0.04 &Z=0.02\\
     & Y=0.25  & Y=0.28 & Y=0.31 & Y=0.33 & Y=0.31 \\
\hline
 3368 &    0.05 &    $-$0.17  & $-$0.12 & $-$0.04 &  $-$0.12\\
 3627 &    0.05 &    $-$0.18  & $-$0.13 & $-$0.05 &   $-$0.09\\
 4182 &    0.04 &    $-$0.13  & $-$0.09 & $-$0.02 &   $-$0.08\\
 4414 &    0.06 &    $-$0.21  & $-$0.15 & $-$0.06 &   $-$0.09\\
 4496A &   0.04 &    $-$0.20  & $-$0.14 & $-$0.06 &   $-$0.09\\
 4527  &   0.07 &    $-$0.21  & $-$0.15 & $-$0.05 &   $-$0.09\\
 4536  &   0.05 &    $-$0.19  & $-$0.14 & $-$0.05 &   $-$0.09\\
 4639  &   0.06 &    $-$0.22  & $-$0.16 & $-$0.07 &   $-$0.10\\
 5253  &   0.04 &    $-$0.10  & $-$0.07 & $-$0.01 &   $-$0.08\\
average & 0.05 &    $-$0.17 & $-$0.13& $-$0.05 &   $-$0.09\\
error   & $\pm$0.02 & $\pm$0.04  & $\pm$0.03  & $\pm$0.02  & $\pm$0.02\\
\hline
\end{tabular}
\end{center}
\end{table}

\begin{table}
\begin{center}
\caption{Predicted corrections (in magnitude) to $\mu^L_{0,0.008}$
as a function of [O/H] with $\Delta Y/\Delta Z$=2.0, 2.5, 3.0, and
3.5.}

\vspace{0.5truecm}

\begin{tabular}{ccccc}
\hline
\hline
      [O/H]&    c(2.0) &      c(2.5)&     c(3.0)&       c(3.5)\\
\hline

      $-$0.700 &   +0.063 &   +0.063 &   +0.064 &   +0.064\\
      $-$0.600 &   +0.040 &   +0.041 &   +0.041 &   +0.042\\
      $-$0.500  &   +0.018 &   +0.018&    +0.019&    +0.019\\
      $-$0.400  & $-$0.019 &  $-$0.014 &  $-$0.009 &  $-$0.005\\
      $-$0.300  & $-$0.087 &  $-$0.077 &  $-$0.068 &  $-$0.059\\
      $-$0.200  & $-$0.146 &  $-$0.129 &  $-$0.113 &  $-$0.097\\
      $-$0.100  & $-$0.193 &  $-$0.167 &  $-$0.141 &  $-$0.117\\
       0  & $-$0.224 &  $-$0.185 &  $-$0.147 &  $-$0.111\\
       +0.100  & $-$0.234 &  $-$0.179 &  $-$0.126 &  $-$0.076\\
       +0.200  & $-$0.219 &  $-$0.143 &  $-$0.072 &  $-$0.005\\
       +0.300  & $-$0.174 &  $-$0.073 &   +0.021 &   +0.108\\
       +0.400  & $-$0.093 &   +0.037 &   +0.157 &   +0.267\\
\hline
\end{tabular}
\end{center}
\end{table}

\clearpage

\figcaption{Theoretical predictions for the instability strip of
fundamental pulsators with the labelled chemical compositions. }

\figcaption{Predicted linear PL$_V$ (upper panel) and
PL$_I$ (lower panel)  relations with the labelled chemical compositions.}

\figcaption{The intrinsic distance modulus $\mu^L_{0,0.008}$ of
fundamental pulsating models with various chemical compositions,
as inferred from the predicted PL$_V$ and PL$_I$ linear relations
with $Z$=0.008, plotted versus the pulsator metal abundance. In
the upper panel, filled and open circles depict the behavior of
short ($P\le$ 10 days) and long period ($P$=10-20 days) pulsators,
respectively, with $\Delta Y/\Delta Z$=2.5, while the triangles
refer to short (filled) and long period (open) pulsators with
$\Delta Y$/$\Delta Z$=4. In the lower panel, the behavior of long
period pulsators (filled circles and triangle) is repeated,
together with the results for very long pulsators ($P\ge$ 20 days,
open circles and triangle). The dotted
lines depict an uncertainty by $\pm$0.1 mag.}

\figcaption{Predicted metallicity correction [eq. (1) and eq.
(2)] to intrinsic distance moduli provided by  LMC-based linear PL
relations, as a function of the pulsator metal abundance and for
selected choices of the $\Delta Y$/$\Delta Z$ ratio. The dotted
lines depict an uncertainty by $\pm$0.1 mag.}

\figcaption{Predicted metallicity correction [eq. (1) and eq.
(2)] for SNIa host galaxies versus the averaged difference between
the distance modulus $\mu^L_{0,Z,Y}$ inferred from $V$ and $I$ PL
linear relations at the various chemical compositions and the result with
$Z$=0.008. 
 Horizontal and vertical error bars are the standard deviation of
the mean and the uncertainty associated to the
predicted metallicity correction, respectively.
(see data in Table 7)}

\figcaption{As in Fig. 4 but as a function of the
oxygen-to-hydrogen ratio (se also Table 8). The vertical arrows
mark the average abundance in the outer and inner fields of M101.
The horizontal arrow depicts the measured difference
(0.16$\pm$0.10 mag) between the LMC-based distance moduli of the
two fields (outer minus inner).}


\begin{references}

\reference{} Alibert, Y., Baraffe, I., Hauschildt, P., \&, Allard, F. 1999, A\&A, 344, 551
\reference{} Baker, N. \&, Kippenhahn, R. 1965, ApJ, 142, 868
\reference{} Bono, G., Caputo, F., Castellani, V., \&, Marconi, M. 1999b, ApJ, 512, 711
\reference{} Bono, G., Caputo, F., Cassisi, S., Marconi, M., Piersanti, L., \&, Tornamb\'e, A. 2000, ApJ, 543, 955
\reference{} Bono, G., Castellani, V.,  \&, Marconi, M. 2000,  ApJ, 529, 293
\reference{} Bono, G., Castellani, V., \&, Marconi, M. 2002, ApJL, 565, 83
\reference{}  Bono, G., Marconi, M., \&, Stellingwerf, R. F. 1999a, ApJS, 122, 167
\reference{}  Bono, G., Marconi, M.,  \&, Stellingwerf, R. F. 2000, A\&A, 360, 245
\reference{} Caputo, F., Marconi, M., \&, Musella, I. 2000a, A\&A, 354, 610
\reference{} Caputo, F., Marconi, M., \&, Musella, I. 2002, ApJ, 566, 833
\reference{} Caputo, F., Marconi, M., Musella, I., \&, Santolamazza, P.  2000b,
A\&A, 359, 1059
\reference{} Caputo, F., Marconi, M., Musella, I., \&, Pont, F.  2001, A\&A, 372, 544
A\&A, 359, 1059
\reference{} Cardelli, J. A., Clayton, G. C., \&, Mathis, J. S. 1989,
ApJ, 345, 245
\reference{} Castelli, F., Gratton, R. G., \&, Kurucz, R. L. 1997a, A\&A, 318, 841
\reference{} Castelli, F., Gratton, R. G., \&, Kurucz, R. L. 1997b, A\&A, 324, 432
\reference{} Chiosi, C., Wood, P. R.   \&, Capitanio, N. 1993, ApJS, 86, 541
\reference{} Christy, R. F. 1962, ApJ, 136, 887
\reference{} Cox, A. N., Stewart, J. N. \&, Lilers, D. D., 1965, ApJS, 11, 1
\reference{} Feast, M. W. \&, Catchpole, R. M. 1997, MNRAS, 286, L1
\reference{} Ferrarese, L. et al. 2000, ApJS, 128, 431
\reference{} Freedman, W. L. 1988, AJ, 96, 1248
\reference{} Freedman, W. L. et al. 1994, ApJ, 435, 31
\reference{} Freedman, W. L. et al. 2001, ApJ, 553, 47
\reference{} Freedman, W. L. \&,  Madore, B. F. 1990, ApJ, 365, 186
\reference{} Iben, I. \&, Renzini, A. 1984, Physics Reports, 105, 6
\reference{} Izotov, Y. I., Thuan, T. X. \&, Lipovetsky, V. A. 1997,
ApJS, 108, 1
\reference{} Kennicutt, R. C. et al. 1998, ApJ, 498, 181
\reference{} Kochanek, C. S. 1997, ApJ, 491, 13
\reference{} Lanoix, P., Paturel, G., \&, Garnier, R. 1999, MNRAS, 308, 969
\reference{} Lequeux, J., Peimbert, M., Rayo, J. F., Serrano, A., \&,
Torres-Peimbert, S. 1979, A\&A, 80, 155
\reference{} Luck, R. E., Moffett, T. J., Barnes, T. G., \&, Gieren, W. P. 1998, AJ, 115, 605
\reference{} Madore, B. F. \&,  Freedman, W. L. 1991, PASP, 103, 933
\reference{} Pagel, B. E. J. \&, Portinari, L. 1998, MNRAS, 298, 747
\reference{} Pagel, B. E. J., Edmunds, M. G., Fosbury, R. A. E., \&,
Webster, B. L. 1978, MNRAS, 184, 569
\reference{} Pagel, B. E. J., Simonson, E. A., Terlevich, R. J., \&, Edmunds,
M. G. 1992, MNRAS, 255, 325
\reference{} Saha, A. et al. 1994, ApJ, 425, 14
\reference{} Saio, H. \&, Gautschy A. 1998, ApJ, 1998, 498, 360
\reference{} Sandage, A. \&, Tammann, G. A. 1968, ApJ,  151, 531
\reference{} Sasselov, D. et al. 1997, A\&A, 324, 471
\reference{} Udalski, A., Soszynski, I., Szymanski, M., Kubiak, M.,
 Pietrzynski, G., Wozniak, P., \&, Zebrun, K. 1999, AcA, 49, 223
\reference{} Unno, W. 1965, PASJ, 17, 205
\reference{} Walker A.,  1999, in Post-Hipparcos Cosmic Candles, eds.
A. Heck, F. Caputo (Dordrecht, Kluwer Academic Publishers), p. 125
\reference{} Zaritsky, D., Kennicutt, R. C., Huchra, J. P. 1994, ApJ 420, 87

\end{references}
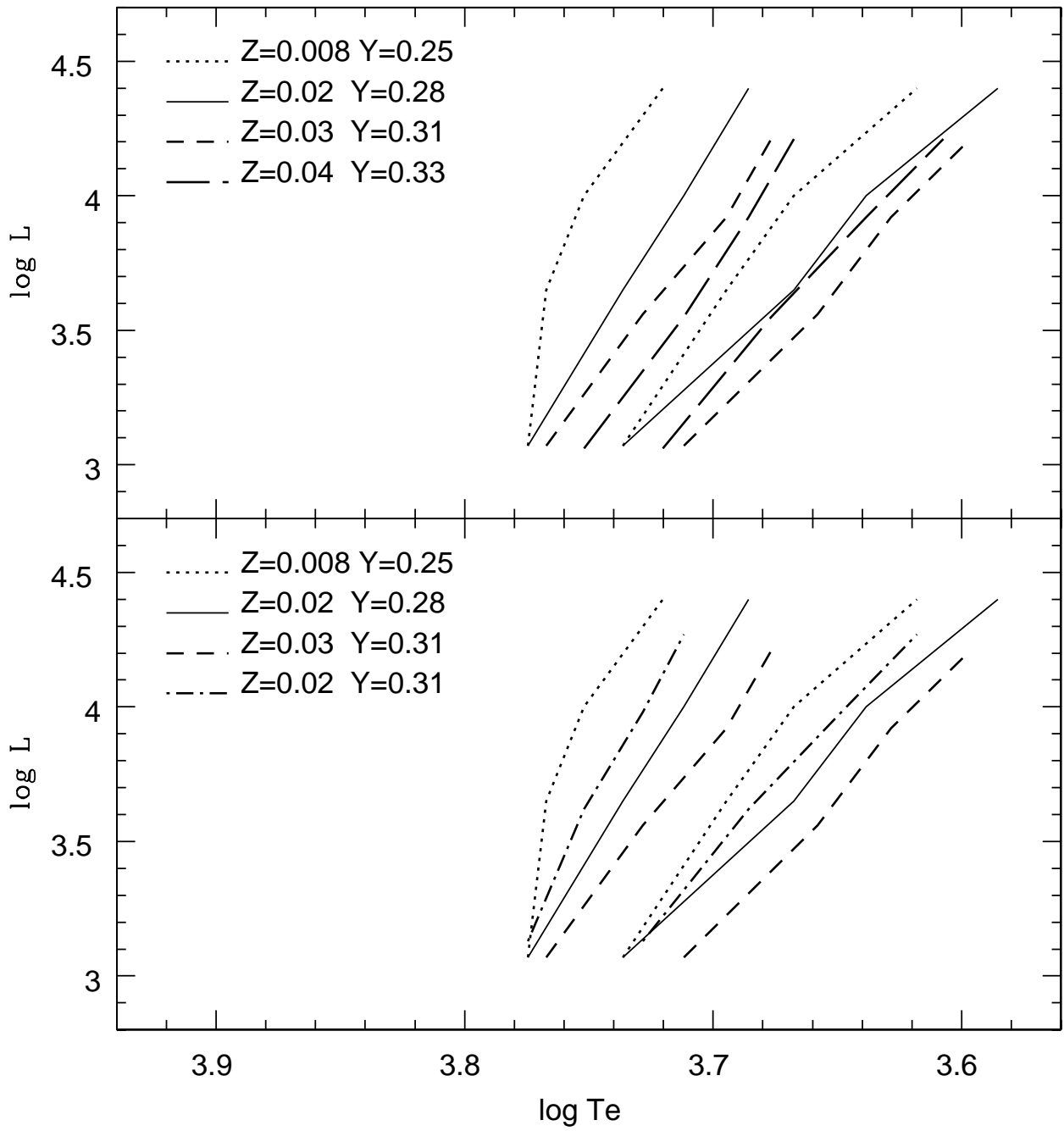
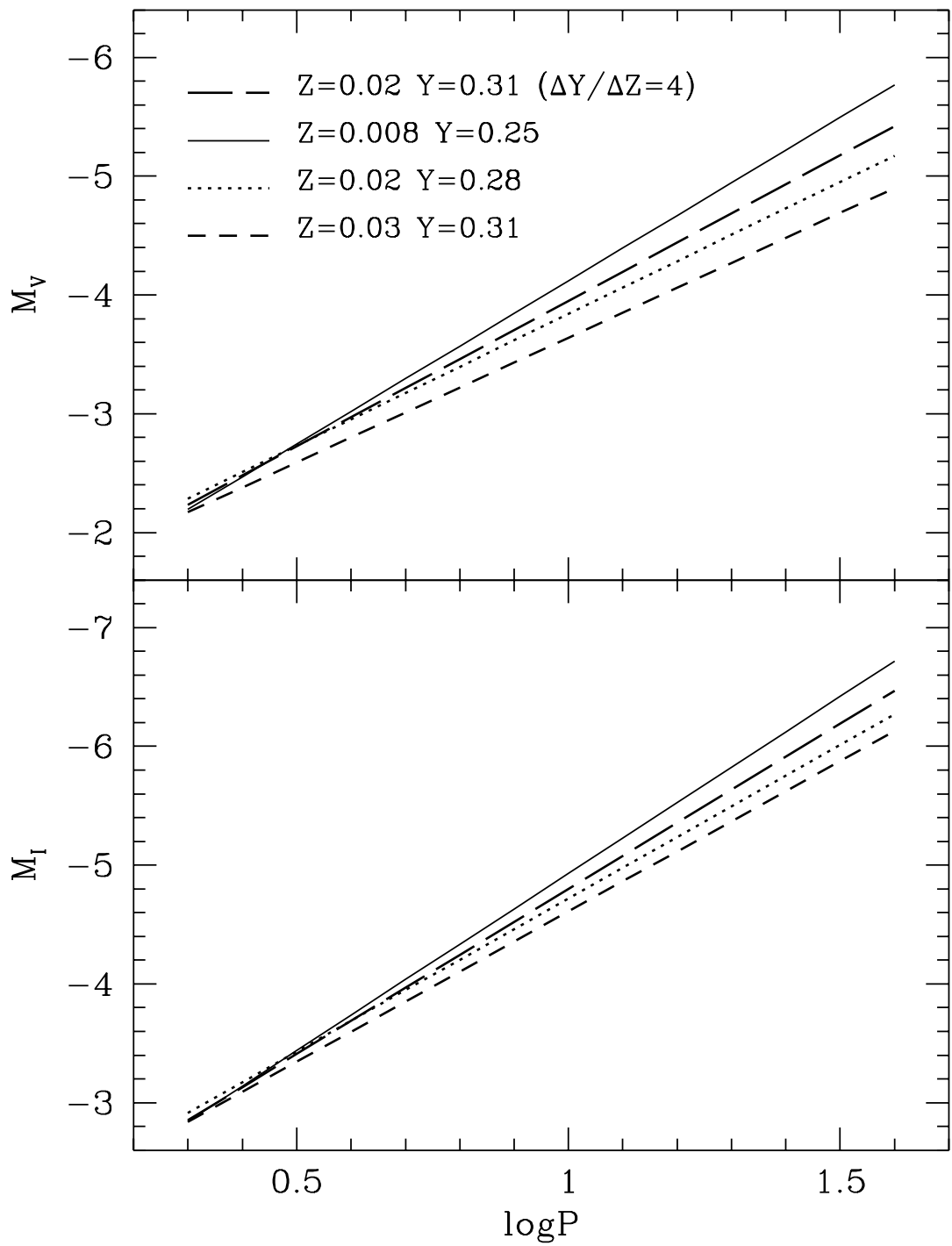
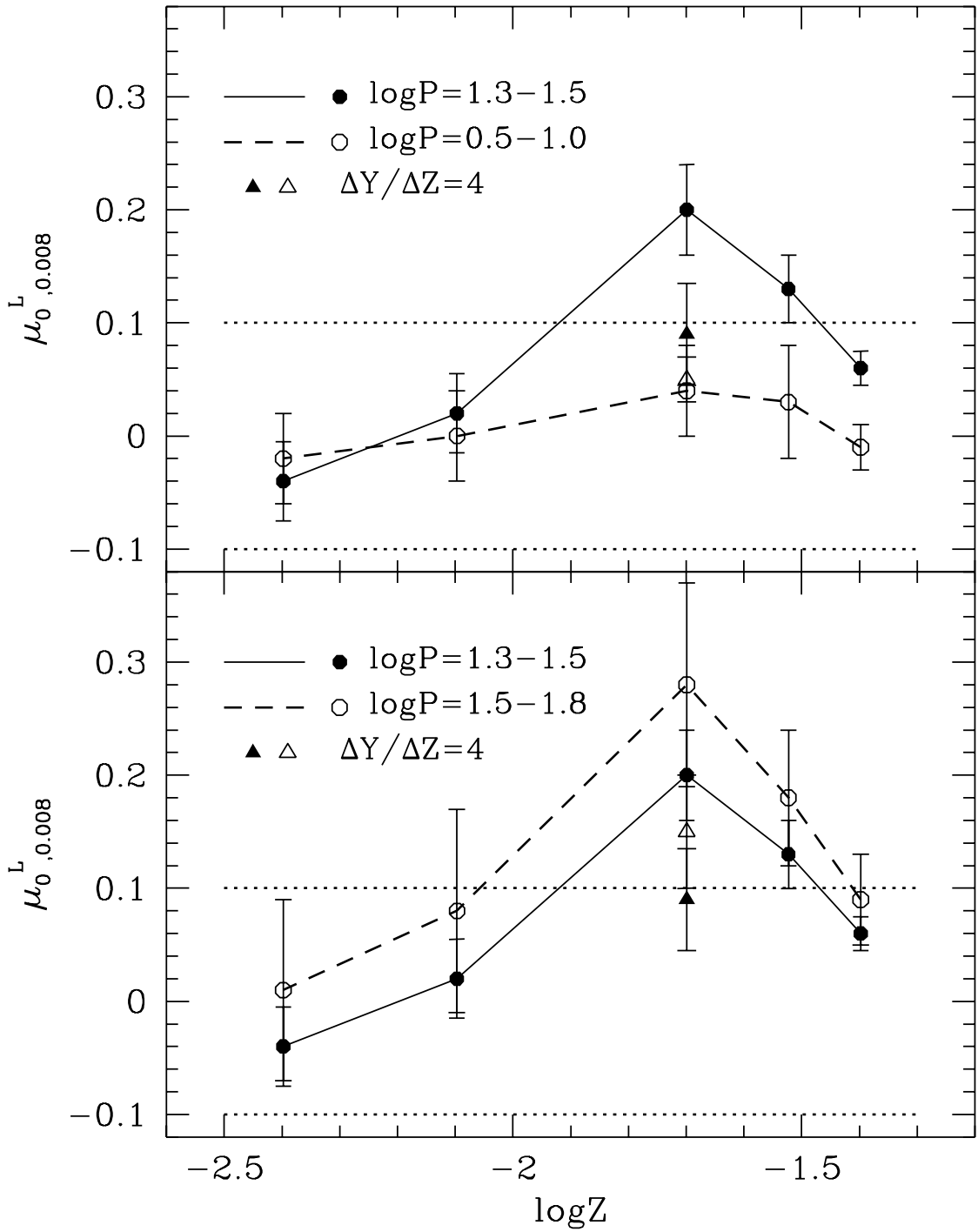
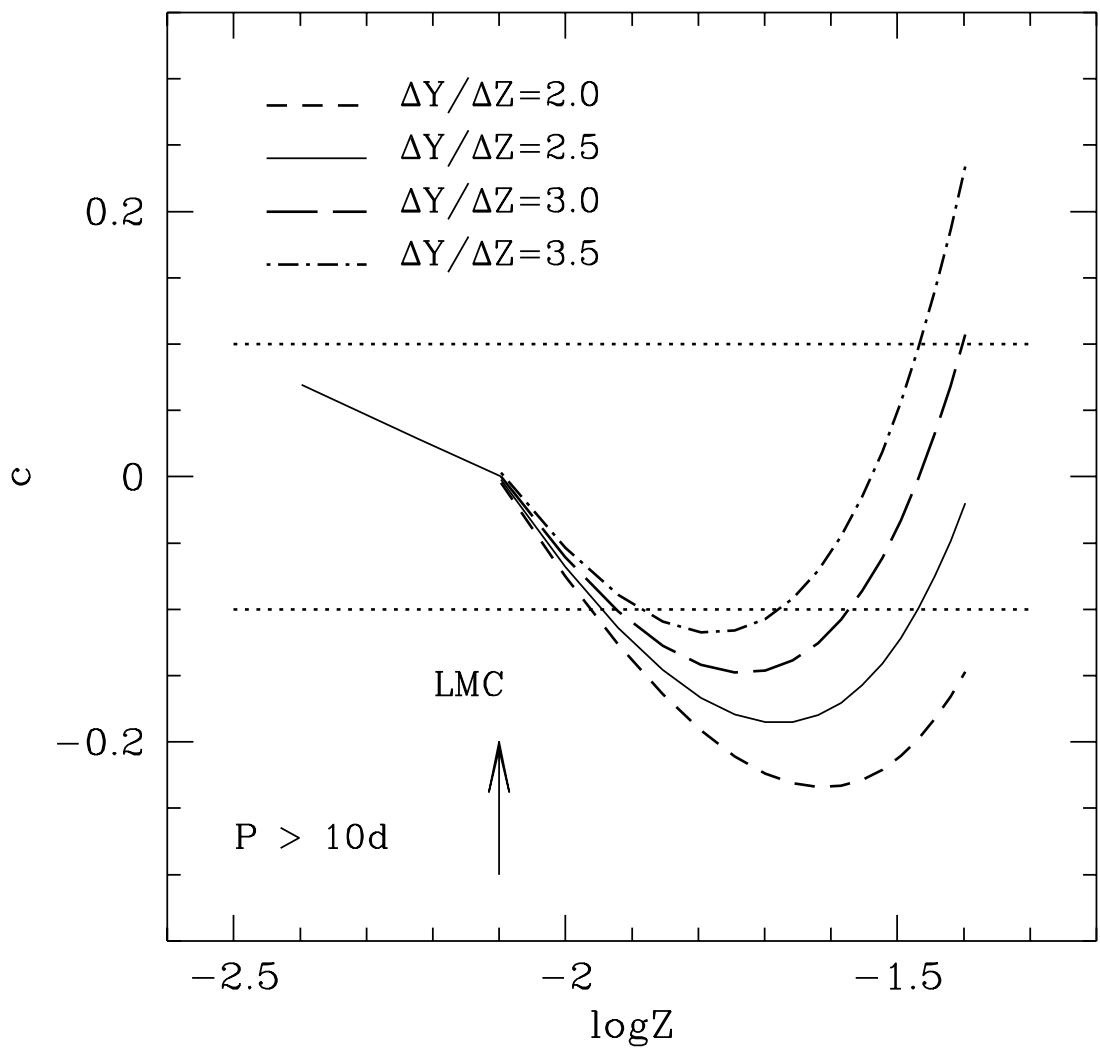
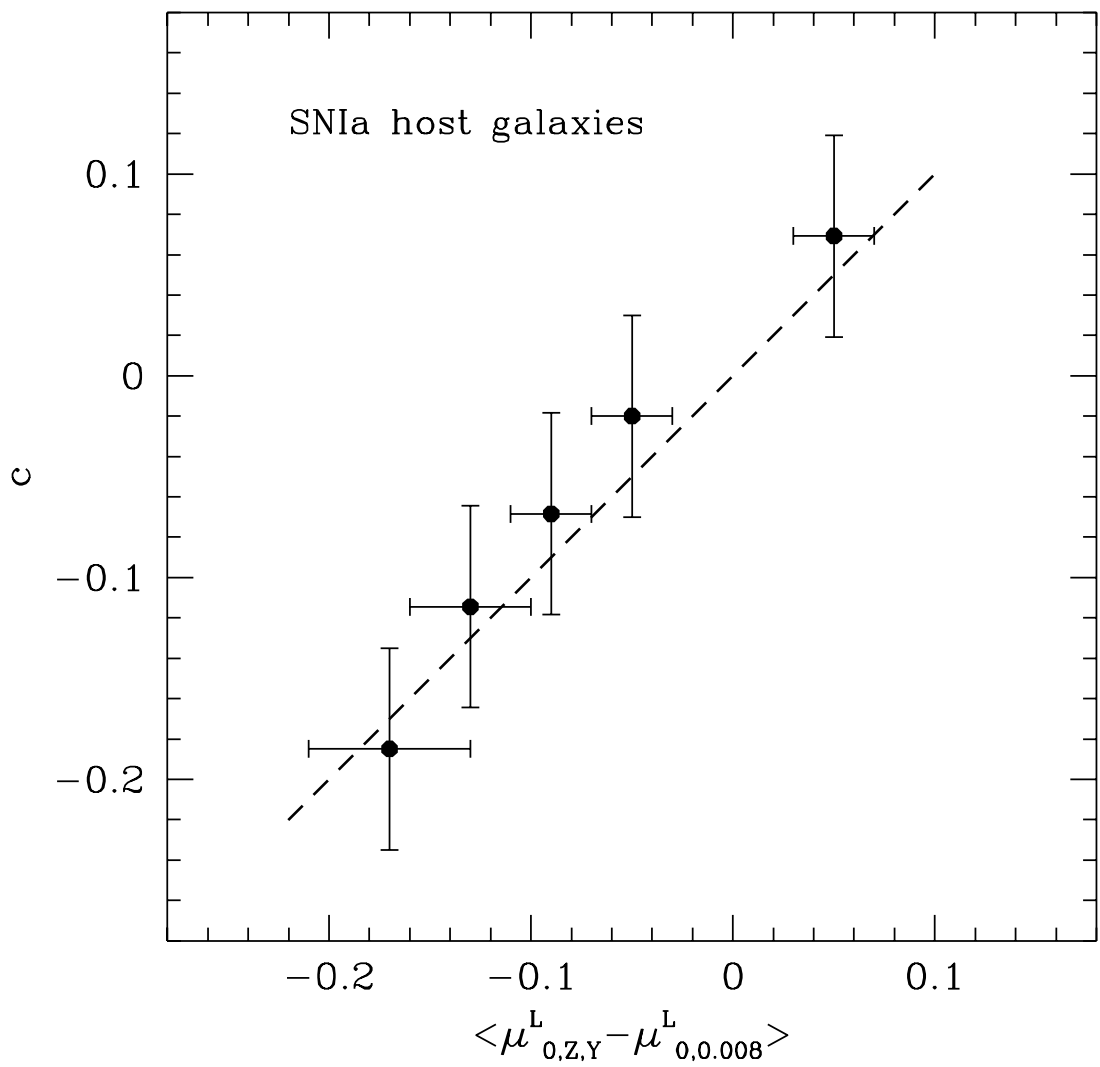
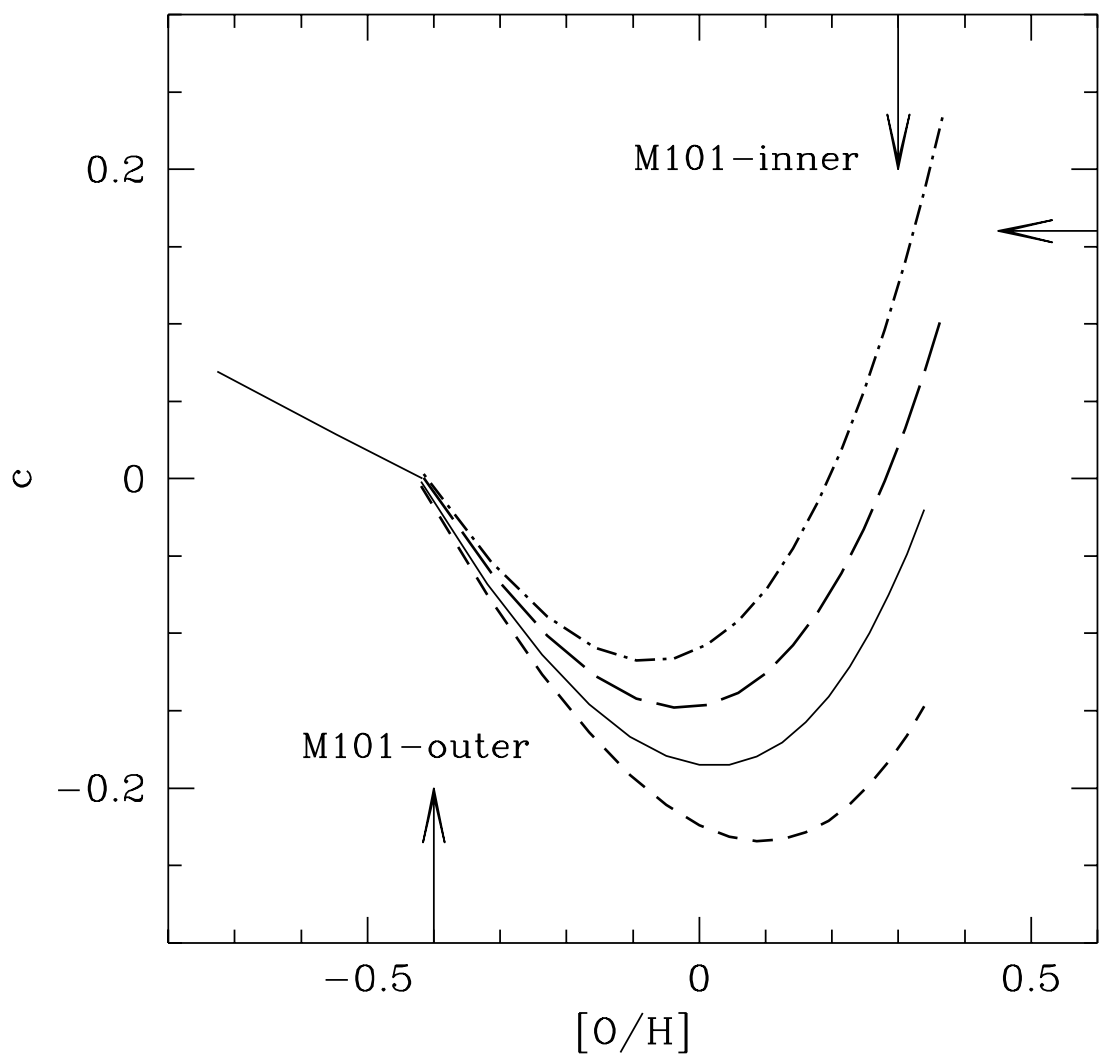
\end{document}